\documentclass [10pt,a4paper]{article}
\usepackage{jheppub}

\usepackage[latin1]{inputenc}
\usepackage{graphicx}
\usepackage{color}
\usepackage{amsfonts,amsthm,amssymb}
\usepackage{bm,bbm}
\usepackage[OT2,OT1]{fontenc}
\usepackage{nicefrac}
\usepackage{axodraw}

\usepackage{verbatim}
\usepackage{units}
\usepackage{mathrsfs}
\usepackage{amsmath}
\usepackage{amssymb}
\usepackage{esint}
\usepackage{slashed}

\newcommand\cyr{%
\renewcommand\rmdefault{wncyr}%
\renewcommand\sfdefault{wncyss}%
\renewcommand\encodingdefault{OT2}%
\normalfont
\selectfont}
\DeclareTextFontCommand{\textcyr}{\cyr}

\def\beq{\begin{equation}}
\def\eeq{\end{equation}}
\newcommand{\be}{\begin{eqnarray}}
\newcommand{\ee}{\end{eqnarray}}

\renewcommand{\texttt}{{}}

\def\bs{\begin{subequations}}
\def\es{\end{subequations}}

\newcommand{\tia}[1]{}

\newcommand{\bea}{\begin{eqnarray}}
\newcommand{\eea}{\end{eqnarray}}
\newcommand{\beas}{\begin{eqnarray*}}
\newcommand{\eeas}{\end{eqnarray*}}
\newcommand{\bal}{\begin{aligned}}
\newcommand{\eal}{\end{aligned}}

\def\({\left(}
\def\){\right)}

\begin{document}

\title{\bf Black supernovae and black holes\\in non-local gravity}

\author[a,b]{Cosimo Bambi}

\author[c]{Daniele Malafarina}

\author[a]{Leonardo Modesto}

\affiliation[a]{Center for Field Theory and Particle Physics and Department of Physics,\\Fudan University, 200433 Shanghai, China}
\affiliation[b]{Theoretical Astrophysics, Eberhard-Karls Universit\"at T\"ubingen, 72076 T\"ubingen, Germany}
\affiliation[c]{Department of Physics, Nazarbayev University, 010000 Astana, Kazakhstan}

\emailAdd{bambi@fudan.edu.cn}
\emailAdd{daniele.malafarina@nu.edu.kz}
\emailAdd{lmodesto@fudan.edu.cn}


\abstract{In a previous paper, we studied the interior solution of a collapsing body in a non-local theory of gravity super-renormalizable at the quantum level. We found that the classical singularity is replaced by a bounce, after which the body starts expanding. A black hole, strictly speaking, never forms. The gravitational collapse does not create an event horizon but only an apparent one for a finite time. In this paper, we solve the equations of motion assuming that the exterior solution is static. With such an assumption, we are able to reconstruct the solution in the whole spacetime, namely in both the exterior and interior regions. Now the gravitational collapse creates an event horizon in a finite comoving time, but the central singularity is approached in an infinite time. We argue that these black holes should be unstable, providing a link between the scenarios with and without black holes. Indeed, we find a non catastrophic ghost-instability of the metric in the exterior region. Interestingly, under certain conditions, the lifetime of our black holes exactly scales as the Hawking evaporation time.}


\maketitle

\section{Introduction}

In Einstein gravity and under a set of physically reasonable assumptions, the complete gravitational collapse of a body creates a spacetime singularity and the final product is a black hole. The simplest example is the Oppenheimer-Snyder~(OS) model, which describes the collapse of a homogeneous and spherically symmetric cloud of dust~\cite{OS}. However, it is often believed that the spacetime singularities created in a collapse are a symptom of the breakdown of the classical theory and they can be removed by quantum gravity effects. Alternatively, we can assume that spacetime singularities are resolved by employing a new action principle for classical gravity. However, the equations of motion of the new theory are typically quite difficult to solve. One can thus attempt to study toy-models, which can hopefully capture the fundamental features of the full theory. With a similar approach, one usually finds that the formation of a singularity is replaced by a bounce, after which the collapsing matter starts expanding~\cite{frolov,Frolov+-1,Frolov+,noi0,noi1,noi2,Tiberio, rovelli, torres, torres2,garay,Siegel,koshe}.

Even in simple models, it is usually quite difficult to find a global solution that covers the whole spacetime. Nevertheless, on the basis of general arguments, we can conclude that there are two plausible scenarios. One possibility is that the bounce generates a baby universe inside the black hole~\cite{baby}. This kind of scenario can generally be obtained analytically with a cut-and-paste technique, in which the singularity is removed and the spacetime is sewed to a new non-singular manifold describing an expanding baby universe. However, such a procedure seems to work only in very simple examples: the matching requires the continuity of the first and of the second fundamental forms across some hypersurface, which is not always possible because of the absence of a sufficient number of free parameters. In the second scenario, a black hole does not form. The gravitational collapse only creates a temporary trapped surface, which looks like an event horizon for a finite time (which may, however, be very long for a far-away observer). Such a possibility has recently attracted a lot of interest because of a paper by Hawking~\cite{hawking}, but actually it was proposed a long time ago by Frolov and Vilkovisky~\cite{frolov,Frolov+-1}, and was recently rediscovered by several groups~\cite{Frolov+,noi0,noi1,noi2,Tiberio, rovelli, torres, torres2,garay}, following different approaches and within different models.

The aim of this paper is to go ahead in the investigation of this topic. Following Ref.~\cite{torres}, we start from a model for the exterior vacuum spacetime. We assume that the exterior metric is static, and we solve our effective equations of motion (EOM) for the non-local gravitational theory. With an ansatz for the interior solution, we are able to do the matching and eventually to obtain a solution for the whole spacetime. The result of this procedure is the formation of a black hole, characterized by a Cauchy internal horizon and an event horizon. More importantly, there is no bounce. The collapsing object approaches a singular state in an infinite time. It seems thus that the properties of the exterior solution, which could in principle be derived by the underlying fundamental theory, play a major role in the fate of the collapse. However, our exterior spacetime metric appears to be unstable because of the presence of a massive ghost. The latter can likely cause the destruction of the black hole, but the timescale is extremely long for a stellar-mass object. We thus argue that, once again, a true event horizon may never be created.

The content of the paper is as follows. In Section~\ref{s-gr}, we briefly review the gravitational collapse of a spherically symmetric cloud in classical general relativity. In Sections~\ref{s-sn}, we summarize the bouncing solutions (black supernovae) in weakly non-local theories of gravity found in~\cite{noi0,noi2}. Moreover, we provide the correct spacetime structure missed in the previous papers. In Section~\ref{s-bh}, we follow the approach of Ref.~\cite{torres} and we construct the interior metric on the base of an external black hole metric \cite{hayward} that captures all the features of the approximate solutions in non-local gravitational theories \cite{mmn}. In Section~\ref{coexistence}, we provide a (in-)stability mechanism to reconcile the contradictory outcome of the previous sections. Indeed, the black hole metric shows a ghost instability which makes the black hole lifetime finite, but very long due to the non-locality scale. Summary and conclusions are reported in Section~\ref{s-c}.

Throughout the paper, we use units in with $c = \hbar = 1$, while we explicitly show Newton's gravitational constant $G_N$.

\section{Gravitational collapse in Einstein gravity \label{s-gr}}

In the case of spherical symmetry, we can always write the line element in the comoving frame as
\be\label{int}
ds^2=-e^{2\nu}dt^2+\frac{R'^2}{Y}dr^2+R^2d\Omega^2 \; ,
\ee
where $d\Omega^2$ represents the metric on the unit 2-sphere. The metric
functions $\nu(r,t)$, $Y(r,t)$, and $R(r,t)$ must be determined by solving
the Einstein equations for a given matter distribution. We note that $R(r,t)$
represents the collapsing areal coordinate, while the comoving radius $r$ is
a coordinate ``attached'' to the collapsing fluid. The energy momentum tensor
in comoving coordinates takes diagonal form and for a matter fluid source
can be written as $T_\mu^\nu={\rm diag}\{-\rho,p_r,p_\theta,p_\theta\}$. With
this set-up, the Einstein equations become
\bea \label{rho}
\rho &= & \frac{F'}{4\pi R^2R'}\; ,  \\ \label{p}
p_r&=&-\frac{\dot{F}}{4\pi R^2\dot{R}} \; ,\\ \label{nu}
\nu'&=&2\frac{p_\theta-p_r}{\rho+p_r}\frac{R'}{R}-\frac{p_r'}{\rho+p_r} \; , \\ \label{Y}
\dot{Y}&=&2\frac{\nu'\dot{R}}{R'}Y \; ,
\eea
where $'$ indicates the derivative with respect to $r$, while $\dot{}$ the one with respect to $t$.
The function $F$ is the Misner-Sharp mass of the system and is defined by (please note that there is a difference of a factor $2 G_N$ in our definition of $F$ with respect our previous papers~\cite{noi0,noi1,noi2})
\be
2 G_N F=R(1-g_{\mu\nu}\nabla^\mu R\nabla^\nu R) \; .
\ee
It is easy to see that $F$
plays the same role as the mass parameter $M_{\rm s}$ in the Schwarzschild metric and represents the amount of gravitating matter within the shell $r$ at the time $t$~\cite{misner}.
Using the metric~(\ref{int}), $F$ can be written as
\be \label{misner}
2 G_N F = R\left(1-Y+e^{-2\nu}\dot{R}^2\right)\; .
\ee

We immediately see that these equations can be considerably simplified if the
matter source satisfies $p_r=p_\theta$ and $p_r'=0$. In this case, we have
$\nu'=0$, from which we get $\nu=\nu(t)$ and, by a suitable redefinition of the time
gauge, we can set $\nu=0$. Eq.~\eqref{Y} becomes $\dot{Y}=0$, which can be
integrated to give $Y=Y(r)=1+f(r)$. A cloud composed of non interacting particles
has $p_r=p_\theta=0$ and satisfies the conditions above. This is the
so called dust collapse and was first investigated in the case of a homogeneous
density distribution in~\cite{OS}. From Eq.~(\ref{p}), we see that in the case of dust
$F=F(r)$ and therefore the amount of matter enclosed within the
shell $r$ is conserved. This means that there is no inflow or outflow of matter at
any radius during the process of collapse. As a consequence, there is no flux
of matter through the boundary of the star as well. Therefore, by setting the
outer boundary of the cloud at the comoving radius $r=r_b$, which corresponds to
the shrinking physical area-radius $R_b(t)=R(r_b,t)$, we see that we can always
perform the matching with an exterior Schwarzschild spacetime with mass
parameter $M_{\rm s}=F(r_b)$~\cite{matching}.

Once we substitute $\nu$ and $Y$ for dust in the definition of the Misner-Sharp
mass given by Eq.~(\ref{misner}), we obtain the equation of motion for the system
\be\label{motion}
\dot{R}=-\sqrt{\frac{2 G_N F}{R}+f} \; .
\ee
The free function $f$ coming from the integration of Eq.~(\ref{Y}) is related
to the initial velocity of the infalling particles. If the cloud had no boundary
and extended to infinity, then the velocity of particles at infinity would be given by
$\lim_{r\rightarrow \infty}f(r)$. This allows us to distinguish three cases. Unbound
collapse happens when particles have positive velocity at infinity. Marginally
bound collapse happens when particles have zero velocity at infinity. Bound
collapse happens when particles reach zero velocity at a finite radius.

There is a gauge degree of freedom given by setting the value of the area-radius $R$ at the
initial time. This sets the initial scale of the system but does not affect the physics
of the collapse. We can choose the initial scaling in such a way that at the initial time
$t_i=0$ we have $R(r,0)=r$ and
introduce a dimensionless scale factor $a(r,t)$ such that $R=ra$. Then the whole
set of the Einstein equations can be rewritten in this gauge once we define two
functions, $\mu(r)$ and $b(r)$, such that
\bea\label{Mscale}
 F= r^3\mu \; , \quad  \label{bscale}
 f=r^2b \; .
\eea
The equation of motion~(\ref{motion}) is immediately rewritten as
\be
\dot{a}=-\sqrt{\frac{2 G_N \mu}{a}+b} \; .
\ee
As a consequence of the above choice, we see that the regularity of the initial data
at the center follows directly from the finiteness of $\mu$ and $b$. This choice makes
also the appearance of the singularity more manifest, since the energy density
becomes
\be
\rho=\frac{3\mu+r\mu'}{4\pi a^2 (a+ra')} \; ,
\ee
which diverges for $a=0$ and is clearly finite at the initial time when $a=1$. As we
can see, the homogeneous dust collapse model is obtained easily by setting
$\mu$ and $b$ to be constant, namely $\mu=\mu_0$ and $b=b_0$. In this case,
marginally bound collapse is simply given by $b_0=0$.  Considering $\mu$ and/or
$b$ as functions of $r$, one gets an inhomogeneous density profile, which
corresponds to the so called Lema\`itre-Tolman-Bondi model (LTB)~\cite{LTB}.
In both the homogeneous and inhomogeneous case, the collapse ends with the
production of a gravitationally strong, shell-focusing singularity. The singularity is
hidden behind the horizon in the OS model, while it may be visible to far-away
observers in the LTB model~\cite{dust}.

\section{Black supernovae \label{s-sn}}

While most of the bouncing solutions are based on
toy-models~\cite{hawking, noi1, rovelli, torres,torres2,garay}, or at best on theories non renormalizable
at the quantum level~\cite{frolov}, in Refs.~\cite{noi0,noi2} we found the bounce
in a family of asymptotically free weakly non-local theories of gravity. These theories
are unitary, super-renormalizable or finite at the quantum level, and there are no
extra degrees of freedom (ghosts or tachyons) expanding around the flat spacetime (for the details, see
Refs.~\cite{noi0,noi2}). The simplest classical Lagrangian for these super-renormaliable theories reads
\cite{kuzmin, Krasnikov, Tombo, Khoury, modesto, modestoLeslaw, universality, Mtheory, Dona}
\be
S_g = \frac{2}{\kappa^2} \int d^4 x \sqrt{|g|}
\Big[R + G_{\mu\nu} \frac{ e^{{\rm H}(-\Box/\Lambda^2)} -1}{\Box}
R^{\mu\nu} \Big] \, ,
\label{theory}
\ee
where $G_{\mu\nu}$ is the Einstein tensor and $\kappa^2 = 32 \pi G_N$.
All the non-polynomiality is in the form factor $\exp {\rm H}(-\Box/\Lambda^2)$, which
must be an entire function. $\Lambda$ is the non-locality or quasi-polynomiality scale.  
The natural value of $\Lambda$ is of order the Planck mass and in this case all
the observational constraints are satisfied. The theory is uniquely specified
once the form factor is fixed, because the latter does not receive any
renormalization: the ultraviolet theory is dominated by the bare action (that is,
the counterterms are negligible). In this class of theories, we only have the graviton
pole. Since $\exp {\rm H}(-\Box/\Lambda)$ is an entire function without zeros or poles in the whole complex plane, 
at perturbative level there are no ghosts and no tachyons
independently of the number of time derivatives present in the action.

Let us now consider the gravitational collapse in the class of theories given by Eq.~(\ref{theory}).
In particular we look for approximate solutions for the interior of a collapsing body.
The scale factor
$a(t)$ is determined via the propagator approach \cite{frolov,noi0,noi2,Calcagni:2013vra,broda} or the linearized equations of motion in the way we are going to describe. We consider a Friedman-Robertson-Walker (FRW) cosmological model since we can easily export the result to the gravitational collapse by inverting the time direction. We start writing the FRW metric
as a flat Minkowski background plus a fluctuation $h_{\mu \nu}$,
\bea
g_{\mu \nu} = \eta_{\mu \nu} + \kappa \, h_{\mu \nu} \, , \quad 
ds^2 = - dt^2 + a(t)^2 dx^i dx^j \delta_{i j} \, ,
\eea
where $\eta_{\mu \nu} = {\rm diag}(-1,1,1,1)$. The conformal scale factor $a(t)$
and the fluctuation $h_{\mu\nu}(t, \vec{x})$ are related by the following
relations:
\bea
&& a^2(t) = 1 - \kappa h(t) \,  , \label{ath}\\
&& h(t=t_0) = 0 \, , \nonumber\\
&& g_{\mu \nu}(t=t_0) = \eta_{\mu\nu} \, , \nonumber\\
&& h_{\mu \nu}(t, \vec{x} ) = - h(t) \, {\rm diag}(0, \delta_{i j} )
\equiv - h(t) \, \mathcal{I}_{\mu \nu} \, .
\label{ah}
\eea
After a gauge transformation, we can rewrite the fluctuation in the usual harmonic
gauge, in which the propagator is evaluated, namely
\bea
&& h_{\mu \nu}(x) \rightarrow h^{\prime}_{\mu \nu}(x) =
h_{\mu \nu}(x)+ \partial_{\mu} \xi_{\nu} + \partial_{\nu} \xi_{\mu} \, ,
\nonumber\\
&& \xi_{\mu}(t) = \frac{3 \kappa}{2} \,
\left( \int_0^{t} h(t') dt',0,0,0  \right) \, .
\eea
The fluctuation in the harmonic gauge reads
\bea
\hspace{-0.4cm}
h^{\prime}_{\mu \nu}(t, \vec{x} ) =
h(t) \, {\rm diag} ( 3, - \delta_{i j} ) \, , \,\, 
h^{\prime \, \mu}_{\mu}(t, \vec{x} ) = - 6 h(t) .
\eea
We can then switch to the standard gravitational ``barred" field
$\bar{h}^{\prime}_{\mu \nu}$ defined by
\be
\bar{h}^{\prime}_{\mu \nu} = {h}^{\prime}_{\mu \nu} - \frac{1}{2} \eta_{\mu \nu} \,
h^{\prime \, \lambda}_{\lambda}
=  2 h(t) \, \mathcal{I}_{\mu \nu} \, ,
\ee
satisfying $\partial^{\mu} \bar{h}^{\prime}_{\mu \nu} = 0$.
The Fourier transform of $\bar{h}^{\prime}_{\mu \nu}$ is
\be
\tilde{\bar{h}}^{\prime}_{\mu \nu}(E, \vec{p})
= 2 \tilde{h}(E) (2 \pi)^3 \delta^3(\vec{p}) \, \mathcal{I}_{\mu \nu} \, .
\label{FThxx}
\ee

For the generic case of a perfect fluid with equation of state $p = \omega \rho$,
the scale factor for the homogeneous and spherically symmetric
gravitational collapse (or cosmological metric) is (for $\omega \neq -1$)
\be
a(t) = \left|  \frac{t}{t_{0}}  \right|^{\frac{2}{3(\omega + 1)}} \, ,
\ee
where now $t=0$ is the time of the formation of the singularity.

 We can thus compute the Fourier transform
$\tilde{h}(E)$ defined in~(\ref{FThxx}). For $\omega \neq -1$, we have
\begin{equation}\label{htilde-general}
\tilde{h}(E)=\frac{2\pi\delta(E)}{\kappa}
+\frac{2\Gamma(\frac{4}{3(\omega+1)}+1)
\mathrm{sin} (\frac{\pi}{2}\frac{4}{3(\omega+1)})}
{\kappa t_0^{\frac{4}{3(\omega+1)}} \left| E \right| ^{\frac{4}{3(\omega+1)}+1}} \, .
\end{equation}
In the case of radiation and dust, we have
\bea
\tilde{h}(E)&=&\frac{2\pi\delta(E)}{\kappa}
+\frac{2}{\kappa t_0 E^2}, \;\;\;
\text{(radiation)} \label{htilde-radi}\\
\tilde{h}(E)&=&\frac{2\pi\delta(E)}{\kappa}+
\frac{4\Gamma(\frac{4}{3})}{\sqrt{3} \kappa t_0^{4/3} \left| E \right| ^{7/3}},
\;\;\; \text{(dust)} \label{htilde-dust} \, .
\eea

Since the theory is asymptotically free, we can get a good approximation
of the solution from the linear EOM of the non-local theory. In particular, given the energy tensor,
we can extract the relation between the Einstein solution and the non-local solution comparing the following
two equations,
\bea
  \Box \bar{h}_{\mu\nu}^{\prime} = 8 \pi G_N  T_{\mu \nu} \, , \qquad 
 e^{H ( \Box ) } \Box \bar{h}_{\mu\nu}^{ \prime \, \rm nl} = 8 \pi G_N T_{\mu \nu} \, ,
\eea
where here $\bar{h}_{\mu\nu}^{\prime}$ is the solution of the linearized Einstein EOM, while
$\bar{h}_{\mu\nu}^{ \prime \, \rm nl}$ is the solution of the linearized non-local EOM.
Therefore, the relation between the two gravitational perturbations is:
\be
 e^{ H ( \Box ) }  \, \bar{h}_{\mu\nu}^{\prime \, \rm nl} = \bar{h}_{\mu\nu}^{\prime} \, .
\ee
In Fourier transform, the above relation reads
\be
 \tilde{\bar{h}}^{\prime\, \rm nl}_{\mu\nu}(k)  = e^{ - H ( k^2) }  \tilde{\bar{h}}_{\mu\nu}^{ \prime }(k) \, ,
 \label{hnl}
\ee
or, for our homogeneous case,
\be
 \tilde{h}^{ \rm nl}(E)  = e^{ - H ( E^2) }  \tilde{h}(E) \, .
 \label{hnle}
\ee
Considering the gravitational collapse for an homogeneous and
spherically symmetric cloud and evaluating the anti-Fourier transform of (\ref{hnl}),
we find the solution for $h(t)$ and then the scale factor $a(t)$ (\ref{ath}).
Everything in this section can be applied to the FRW cosmology as well as to the gravitational collapse.
The solution for the gravitational collapse scenario is obtained by replacing $t$ with $-t + t_0$.
For instance, in the
radiation case and for the form factor $\exp( - \Box/\Lambda^2)$, the result is~\cite{noi0}
\be\label{a1}
a^2(t) = \frac{ 2 e^{-\frac{1}{4} \Lambda^2 (t-t_0)^2}}{\Lambda \sqrt{\pi } \,  t_0} +
\frac{ (t_0 -t) \, \text{erf} \left(\frac{\Lambda (t_0 - t)}{2}\right)}{ t_0}  \, ,
\ee
where ${\rm erf}(z) = 2\int_0^z \exp(-t^2) dt/\sqrt{\pi}$. The classical singularity is
now replaced by a bounce at $t=t_0$, after which the cloud starts expanding
(hence the name black supernova). For dust, we find a very similar solution \cite{noi0}. 
The resulting profile for $a(t)$ is slightly different if we consider consistent form factor in Minkowski signature \cite{FrolovEven},
namely $\exp (\Box^{N})$, where $N$ is an even integer. 
It is a general feature of these theories that the gravitational interaction is switched off at high energies, namely the theories are
asymptotically free.
In our framework, the asymptotic
freedom is due to a higher derivative form factor, which makes gravity repulsive
at very small distances. In terms of an effective picture in which gravity is supposed
to be described by the Einstein-Hilbert theory and new physics is absorbed into the matter
sector, the bounce comes from the conservation of the (effective) energy-momentum
tensor: matter is transformed into a state with $\rho_{\rm eff} + p_{\rm eff} < 0$,
which is unstable and therefore the bounce is the only available possibility.

The bounce seems thus to be unavoidable in this class of theories. If we exclude
the possibility of the creation of a baby universe, motivated by the problems mentioned
in the introduction, a black hole, in the strict mathematical sense of the definition, never forms. Gravitational collapse only produces a trapped surface lasting for a finite time.
No Cauchy and event horizon are formed. Since an apparent
horizon cannot be destroyed from the inside, at least if we do not invoke exotic
mechanisms like super-luminal motion, it must be destroyed from the outside. We
thus argue that the solution outside the horizon cannot be static but must belong to the radiating Vaidya family.
We can think of it as an effective negative energy flux destroying the horizon from
the exterior. For a large black hole, we do not expect significant deviations from
standard general relativity at the horizon (the value of scalar quantities like the
Kretschmann invariant is much smaller than the Planck scale) and therefore the
process is expected to be very slow. In other words, we recover the classical picture
of an almost classical black hole and we can realize that the object is not a black
hole only if the observation of a far-away observer lasts for a very long time.

\begin{figure}
\begin{center}
\boxed{\includegraphics[height=5.2cm]{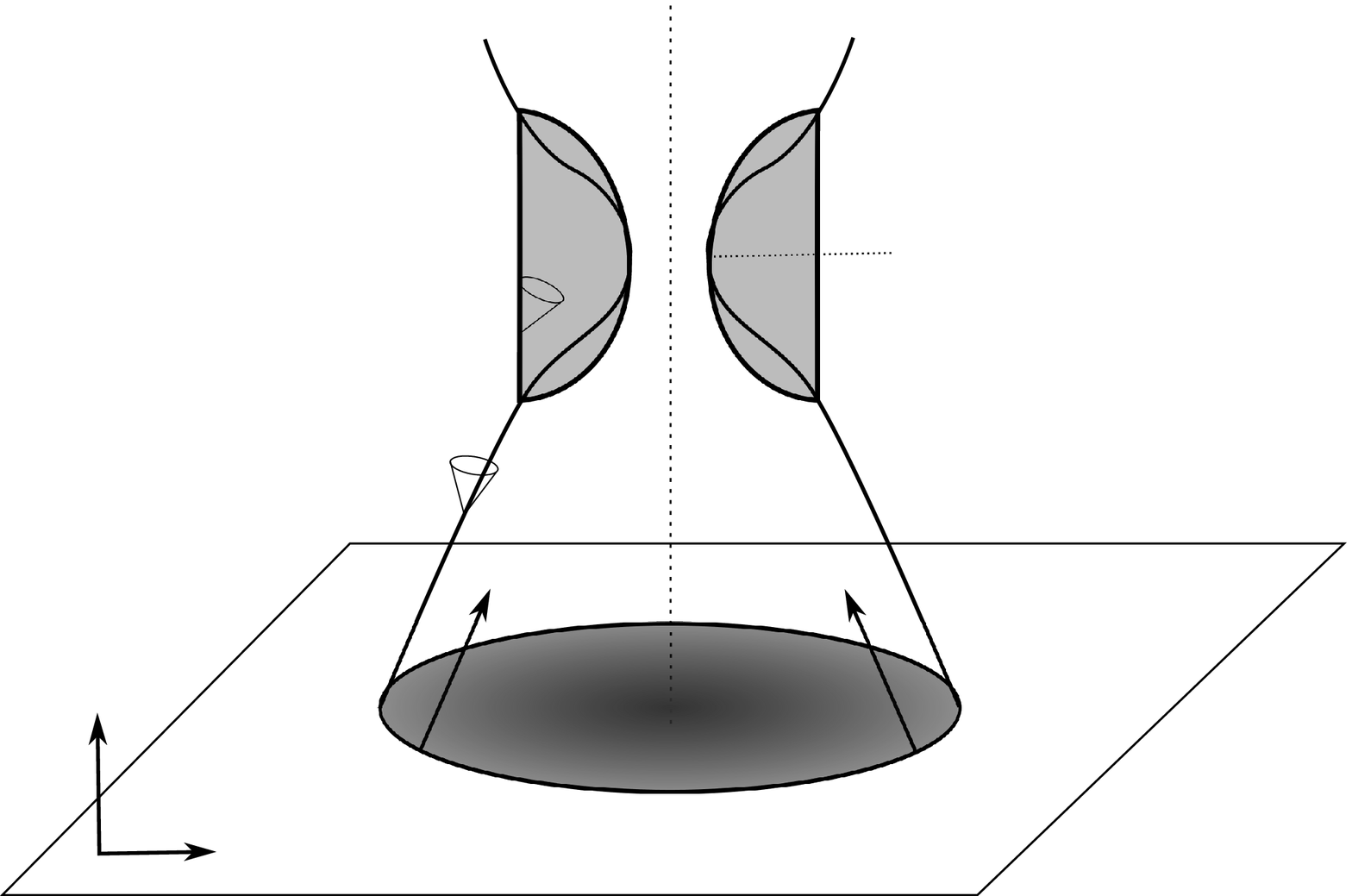} 
\hspace{-1cm}
\includegraphics[height=5.2cm]{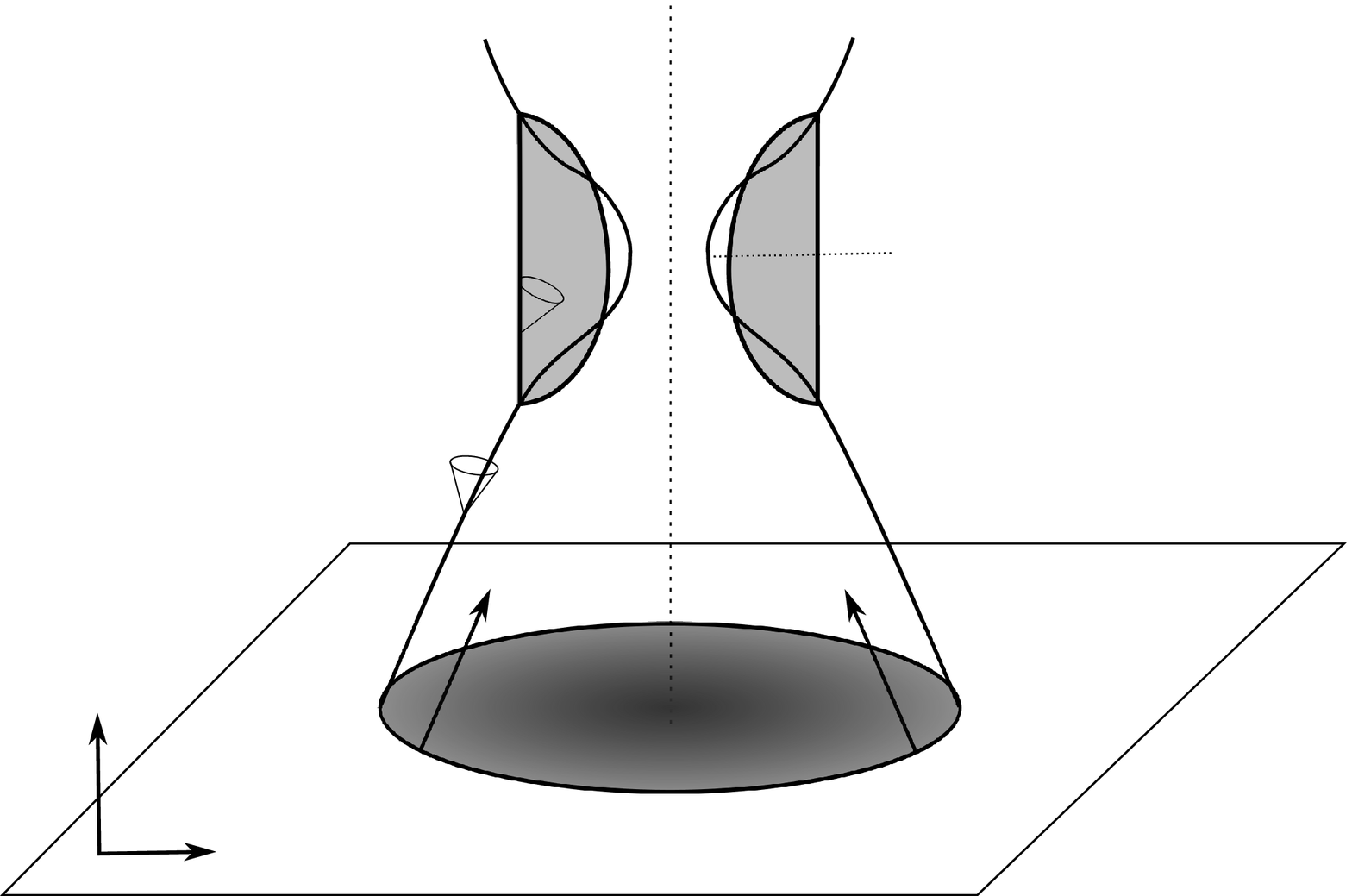}}
\end{center}
\caption{Finkelstein diagram of the black supernova scenario. The two panels differ for the 
position of the Cauchy horizon with respect to the boundary of the cloud. 
However, the spacetime structure of the gravitational collapse has a universal feature characterize by the formation of a trapped surface without any final black hole state. 
 \label{fig1}}
\end{figure}

\begin{figure}
\begin{center}
\boxed{\includegraphics[height=11cm]{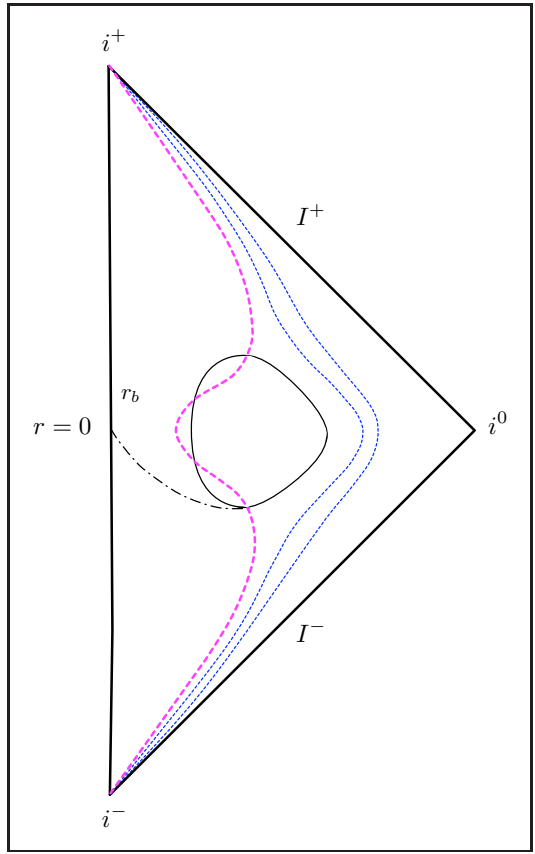} }
\end{center}
\caption{Penrose diagram of the black supernova scenario. There is a single
trapped surface, which behaves for a finite time first as a black hole and then as a
white hole. See the text for more details. \label{fig2}}
\end{figure}

In summary, with the approach employed in Ref.~\cite{noi0,noi2} we start with a
well-defined and consistent theory of gravity for the interior solution and we find
that the bounce is unavoidable. On this basis, we can guess the exterior behavior.
Fig.~\ref{fig1} shows the Finkelstein diagram of the collapse. Fig.~\ref{fig2}
shows instead the corresponding Penrose diagram. We note that the latter corrects
current diagrams presented in the literature. There is more likely only one trapped
surface (not two), because gravity is switched off only inside the cloud of matter.
The apparent horizon propagating inward from the cloud surface 
may either coincide with
the cloud surface at the moment of the bounce (left panel in Fig.~\ref{fig1}) or be in the exterior region (right panel). The actual situation may depend on the gravity theory. In our case we do not know because we are only able to solve the interior solution, so we cannot make predictions about the exterior region. The right panel in Fig.~\ref{fig1} may be motivated by the fact that the static black hole solutions in these theories have indeed an internal Cauchy horizon~\cite{noi2}.
For a finite observational time, the trapped surface first behaves
as a black hole (left bottom side of the trapped surface in Fig.~\ref{fig2}) and then
as a while hole (left top side)~\cite{rovelli}.

\section{Black holes  \label{s-bh}}

In this section, we employ a semi-classical picture
in which deviations from the classical theory are encoded in an effective Newton
gravitational constant. $G_N$ is replaced by a function $G$ of the radial coordinate, which is
used to reproduce the effects of (\ref{theory}) or a generic quantum effective action for gravity~\cite{torres,tavakoli}. 
To this aim we start from the exterior solution and we reconstruct the interior.

\subsection{Exterior solution}

As done
in~\cite{torres, torres2}, we assume that the exterior metric is a generalization of the
classical Schwarzschild solution. The line element can be
written as
\bea\label{ext}
ds^2 = - \left( 1 - \frac{2 G( x ) M_{\rm s}}{x} \right) dt^2 
 + \left( 1 - \frac{2 G( x ) M_{\rm s}}{x} \right)^{-1} dx^2 + x^2 d \Omega^2 \; ,
\eea
where $x$ is the radial coordinate in the exterior spacetime. In super-renormalizable/finite
theories of gravity, spherically symmetric exact black hole solutions can be written
in this form~\cite{noi2,mmn}. Notice the following key point: we are assuming that
the exterior vacuum metric is static, as it is true in general relativity thanks to the
Birkhoff theorem. A prototype of $G(x)$ that captures all the important and universal
features in these theories has the following form
\be
G(x) = \frac{x^3 G_N}{x^3 + L^3} \; ,
\label{prototype}
\ee
where $L$ is a new scale and it is natural to expect it to be of order the Planck length,
namely $L \approx L_{Pl} = G_N^{1/2}$. Of course, Eq.~(\ref{ext}) is not a vacuum
solution of the Einstein equations. If we impose the latter, we find an effective, or
``unphysical'', matter source for the spacetime in the form of an energy-momentum
tensor for a fluid with effective density and pressures given by
\bea
\rho^{\rm ext}=-p^{\rm ext}_r=\frac{M_{\rm s}G_{,x}}{4\pi G_Nx^2} \; , \quad 
p^{\rm ext}_\theta= -\frac{M_{\rm s}G_{,xx}}{8\pi G_Nx} \; .
\eea
New physics is encoded in $G(x)$, but one could have equivalently
absorbed everything in a variable mass parameter $M(x)$, as done in~\cite{noi2,mmn}.
In the next subsection, the line element in~(\ref{ext}) will be matched to a suitable
interior in the form of~(\ref{int}) through a 3-dimensional hypersurface
$\Sigma$ describing the boundary of the collapsing cloud.

\subsection{Interior solution}

The use of a non-constant $G$ in the interior will affect the energy-momentum
tensor by introducing some effective terms in the density and in the pressures. If
$\Sigma$ is the comoving boundary hypersurface, then continuity of
$g_{\theta\theta}$ and $g_{\phi\phi}$ implies that $R(r, t)|_{\Sigma}=R(r_b, t) = x_b(\tau)$.
We can then take the function $G(x)$ from the exterior and obtain the
corresponding $G(R)$ in the interior through the matching conditions.
Standard matching conditions imply continuity of the first and second
fundamental forms across $\Sigma$~\cite{matching}, namely the metric
coefficients on the induced metric and the rate of change of the unit normal
to $\Sigma$ must be the same on both sides. With the exterior metric
given in Eq.~(\ref{ext}), the matching conditions across $\Sigma$ imply that the
density and the pressures in the interior take the form
\bea
\rho= \frac{G(R)F'}{4\pi G_NR^2R'}+\frac{FG_{,R}}{4\pi G_NR^2} \; , \quad 
p_r = - \frac{FG_{,R}}{4\pi G_NR^2} \; , \quad 
p_\theta=  -\frac{FG_{,RR}}{8\pi G_NR} -\frac{F'G_{,R}}{8\pi G_NRR'}\; ,
\eea
which reduce to the usual Einstein equations for dust in the case $G = G_N$
is constant. From these equations and Eq.~(\ref{nu}), we find
that $\nu'=0$,
and therefore the metric in the interior region still
satisfies the same condition as the classical dust case. The line element
can then be taken as
\be\label{LTB}
ds^2 = - d\tau^2 + \frac{R'(r, \tau)^2}{1+ f( r )} dr^2 + R^2(r, \tau) d \Omega^2 \; .
\ee
This is the usual LTB spacetime describing the collapse
of a dust cloud, where now the energy-momentum tensor is the sum of the
classical dust energy momentum-tensor and an effective contribution coming
from the fact that $G$ is not constant. The equation of motion for the system
becomes
\be
\dot{R}^2 = \frac{ 2 G( R ) F( r )}{R} + f( r ) \; .
\label{equation}
\ee

At this point, we have to specify the expression of $G(R)$ for the interior. As an
example, for the sake of simplicity we consider a modified Hayward metric~\cite{hayward} that gives an 
equation (\ref{equation}) independent on the coordinate $r$, namely
\be
G(R) = \frac{R^3 G_N}{R^3 + G_N F(r)  L_{\rm Pl}^2 } \; .
\label{prototype4}
\ee
In the simplest case of a homogeneous cloud, $F(r) = \mu_0 r^3$ with $\mu_0$ constant. Therefore
\be
G(a) = \frac{a^3 G_N}{a^3 +  G_N \mu_0  L_{\rm Pl}^2 } \,  ,
\label{prototype5}
\ee
which is independent on the radial coordinate $r$. With the further assumption
of marginally bound collapse, namely $f=0$, Eq.~(\ref{equation}) becomes
\be
\frac{\dot{a}^2}{a^2} = \frac{2 G_N \mu_0 }{a^3 + G_N \mu_0 L_{\rm Pl}^2} \; .
\label{simple}
\ee
Eq.~(\ref{simple}) can be integrate from $a$ to $1$, 
namely
\bea
\hspace{-0.7cm}
\sqrt{2 G_N \mu_0} \,  t = 
\frac{2}{3} \left(-\sqrt{a^3+c}+\sqrt{c} \tanh
   ^{-1}\left(\sqrt{\frac{a^3+c}{c}}\right)\right.  
   \left.   +\sqrt{c+1}-\sqrt{c} \tanh
   ^{-1}\left(\sqrt{\frac{1}{c}+1}\right)\right) \, , 
\eea
where $c = G_N \mu_0 L_{\rm Pl}^2$.
The classical solution can be recovered in the limit $c\rightarrow 0$
\be
\sqrt{2 G_N \mu_0} \,  t = \frac{2}{3} \left(1- a^{3/2}\right) .
\label{GRc}
\ee

\begin{figure}
\begin{center}
\includegraphics[height=8cm]{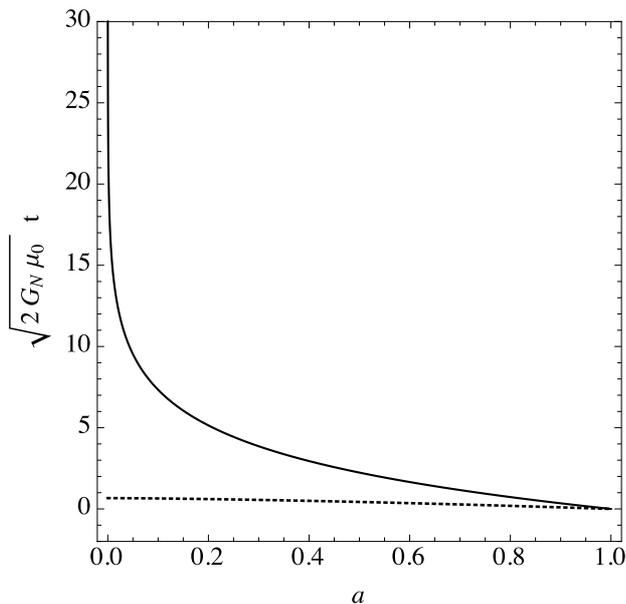}
\end{center}
\caption{Behavior of the scale factor $a(t)$ in the black hole scenario. The
singular state with $a=0$ is approached in an infinite time. 
Therefore, a black hole forms presenting a Cauchy horizon and an event horizon. 
\label{fig3}}
\end{figure}

\begin{figure}
\begin{center}
\boxed{
\includegraphics[height=5.5cm]{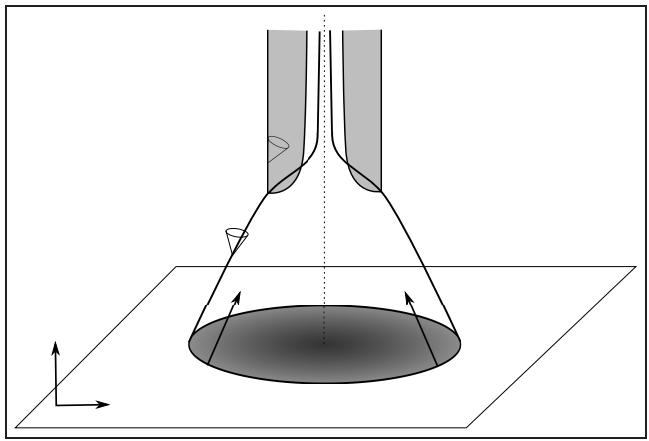}
}
\end{center}
\caption{Finkelstein diagram of the black hole scenario. See the text for more
details. \label{fig4}}
\end{figure}

The behavior of the scale factor is shown in Fig.~\ref{fig3} (solid line). The
singular state $a=0$ is approached in an infinite time. For comparison,
Fig.~\ref{fig3} also shows the case of general relativity (dashed line) whose analytic expression is given in (\ref{GRc}). In the GR case $a=0$ is reached in a finite time. The Finkelstein diagram of this collapse
is shown in Fig.~\ref{fig4}. It is clear that in this scenario we have a real black
hole with a Cauchy horizon and an event horizon.

\subsection{From in to out making use of the boundary conditions}

The gravitational collapse and the cosmological solutions previously obtained in the asymptotic free limit of the weakly non-local theories are all consistent with
a general effective FRW equation for the interior matter bouncing. 
This is a universal property of super-renormalizable asymptotically free gravitational theories
including the recent proposed Lee-Wick gravities \cite{ShapiroComplex, MeShapiro, MeLW}. 
The simplest effective FRW equation compatible with the general feature discussed in Section~\ref{s-sn} reads
\bea
H^2 = \frac{ \dot{a}^2 }{a^2} = \frac{8 \pi G_N}{3} \left( 1-\frac{\rho}{\rho_c} \right) \quad   \, {\rm or}
\quad 
 \frac{ \dot{a}^2  }{2} = \frac{4 \pi G_N}{3} \rho_0 \left( \frac{a^3 - a_c^3}{a^3} \right) \frac{1}{a} \; .
 \label{guess}
\eea
Here we only consider the homogeneous interior. Applying again the ``Torres" procedure
to reconstruct the metric in the vacuum from the metric in the matter region,
we get the exterior spacetime imposing that the boundary conditions of the previous sections are satisfied.
Comparing the interior FRW equation (\ref{guess}) with 
(\ref{equation}), we can derive the effective scaling of the Newton constant with the radial coordinate, namely
\be
G( x ) = \frac{x^3 - l_{\Lambda}^3}{x^3} \,  ,
\label{Gx2}
\ee
where $x$ is the radial coordinate.
The exterior Schwarzschild spacetime is again (\ref{ext}).
The metric is singular in $x=0$, but our derivation is correct only for
$x \geqslant x_{\rm bounce} = l_{\Lambda}$, and $x_{\rm bounce}$ is a finite positive value.
Therefore the metric (\ref{ext}) with (\ref{Gx2}) is only valid for $x  \geqslant l_{\Lambda}$.
The Cauchy and event horizons, if any, are located where the function $g_{00}( r )$ vanish.
For different values of the mass $M$ we can have two roots, two coincident roots, or zero roots.
Therefore, we here provide a justification for the diagrams in Section~\ref{s-sn} that are only correct 
whether the metric in the external region present a Cauchy horizon. Nevertheless, this is the spacetime 
structure of any astrophysical object with $M \gg M_{\rm Pl}$ and then the 
metric in this subsection, by construction, is compatible with the internal matter bounce. 
For completion, the Kretschmann invariant is
\be
R_{\mu\nu\rho \sigma} R^{\mu\nu\rho \sigma}= \frac{48 G_N^2 M^2 \left(39 l_{\Lambda}^6-10 l_{\Lambda}^3 r^3+r^6\right)}{r^{12}} \, .
\ee

\section{Coexistence of the two scenarios and Hawking evaporation}\label{coexistence}

The bouncing (black supernova) and the non-bouncing (singularity free
black hole) solutions seem two different scenarios emerging from
the same theory. In our class of weakly non-local gravities~(\ref{theory})
and in many other frameworks~\cite{frolov,noi1,rovelli,torres, torres2,garay}, the
bounce appears to be unavoidable. However, we do not have the metric
of the whole spacetime under control. If we make the reasonable 
assumption that the exterior vacuum solution is static,
we end up with a regular black hole. The final product of the
collapse would thus depend on whether we reconstruct the external spacetime 
(imposing the boundary conditions for the continuity of the metric and its first derivatives)
from the approximate solution inside the matter (Section~\ref{s-sn}) or the matter interior spacetime from 
the static metric outside the collapsing body. 
While at the moment we cannot completely exclude the coexistence of both the dynamics, we would like to provide another possibility.

In this section we provide a mechanism to reconcile the two scenarios 
based on the stability analysis of the spacetime outside the matter region.

As we have already pointed out in Ref.~\cite{noi2},
it is quite mysterious that in our class of weakly non-local theories of gravity~(\ref{theory})
we can find the bouncing solution when we consider the gravitational
collapse of a spherically symmetric cloud of matter and, on the other hand, 
regular black hole (approximate) solutions when we consider the static case. 
It is possible that all these regular black hole spacetimes are not stable and that their instability
provides a link between the bouncing and non-bouncing scenarios.

The black hole solutions are indeed characterized by a de~Sitter core,
in which the effective cosmological constant is proportional to the mass
of the collapsing object~\cite{noi2}.
From an analysis of the propagator, we can infer that there is a ghost-like pole,
namely the spacetime is unstable. We can thus expect that the black
hole decays into another black hole state with a de-Sitter core with a
smaller effective cosmological constant in one or more steps through
metastable configurations. The process should end when the effective
cosmological constant is of the order of our non-local scale $\Lambda$,
likely close to the Planck mass $M_{\rm Pl}$ if we identify the two scales in the theory.
A solution with a de~Sitter core proportional to $M_{\rm Pl}$ is not a black
hole but a ``particle'' with a sub-Planck mass and without Cauchy and
event horizons. Even if we do not know the intermedia states, the stability
analysis may suggest that the black supernova and regular black hole
scenarios are two faces of the same coin.
In this way we also provide a reasonable justification for the well known instability 
of the Cauchy horizon. In our picture, the Cauchy horizon is just a sector of the close trapped surface,
which of course do not extend to infinity. 
In all the approximate black hole solutions 
studied in the past \cite{noi2,mmn}, three possible different spacetime structures
were presented depending on the value of the mass: with two event horizons, with two coincident horizons (extremal black hole case), and without any event horizon (Planck mass particle).
However, the correct way to interpret such spacetimes is not as unstable black hole because of the
Cauchy horizon, but as different phases of the collapse and bounce (black supernova).

\begin{figure}
\begin{center}
\includegraphics[height=7cm]{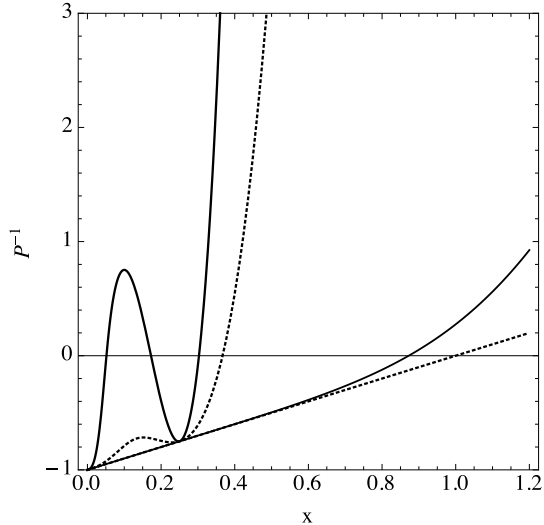}
\end{center}
\caption{Plot of the inverse propagator (\ref{deSitterProp}) for $8 H^2/\Lambda^2 = 1,10,25$. The lowest dashed curve corresponds to the local two derivative case, namely $\Lambda \rightarrow + \infty$ and 
$P^{-1} \propto x-1$. 
Here we used the following form factor: 
${\rm H}(z) = \frac{1}{2} \left(\log \left(z^4\right)+\Gamma \left(0,z^4\right)+\gamma_E \right)$.
\label{figghost}}
\end{figure}

Let us now expand on the ghost-instability. 
While a spacetime with a ghost-instability compatible with the optical theorem in general does not exist at
all \cite{Cline}, because its decay time is not small but exactly zero, this is not true for
weakly non-local theories~\cite{vilenkin}, and our class of
theories~(\ref{theory}) belongs to this group. 
It is crucial to notice that 
the singularity-free black hole metrics always show a de Sitter core with a huge effective cosmological constant, namely
\be
\Lambda_{\rm eff} \approx M G_N \Lambda^3 \, ,
\ee
where $M$ is the mass of the body. 
Therefore, we can easily calculate the second variation of the action (\ref{theory}) for the tensor perturbations around the de Sitter spacetime, namely 
\be
g_{\mu\nu} = \bar{g}_{\mu\nu} + h_{\mu \nu}
\ee
where $\bar{g}_{\mu\nu}$ is the de Sitter metric. Here, we also use the parametrization
\be
ds^2 = - dt^2 + \exp ( 2 H t) d\vec{x}^2 \, ,
\ee 
where $8 H^2 = 8 \Lambda_{\rm eff}/3$.
Moreover, the non-vanishing components for the tensor perturbations are purely spatial, 
$h^0_\mu$= 0, and satisfy the usual transverse and traceless conditions: 
$h^i_i=0$, $\partial_i h^i_j=0$.
 This computation was done for the first time in the paper \cite{khoury} without introducing any cosmological constant in the action. The final result for the variation of the action reads
 \bea
 && \hspace{-0.4cm} 
  \delta S_g  =  2 \kappa_4^{-2} \int d^4 x \sqrt{| \bar{g}|} \, 
 \frac{1}{4} h_{ij} 
  [ (\Box - 8 H^2) + (\Box - 2 H^2) \gamma(\Box) (\Box - 2 H^2) ] h^{ij } 
   \nonumber  , \\
  && \hspace{-0.4cm} 
  \gamma(\Box) = \frac{e^{{\rm H}( - \Box/\Lambda^2)} - 1}{\Box} \, . 
 \label{deSitterPer}
 \eea
From the definition $\Box = - 8 H^2 q^2 = 8 H^2 x$ (here we introduced a basis of eigenfunctions 
$h_{i j}^{(q)}$ for the $\Box$ operator, with dimensionless momentum eigenvalues $-q^2$), the inverse propagator is
\be
\frac{P^{-1}(x)}{4 H^2 \kappa_4^{-2} } = x-1 + \left( x - \frac{1}{4} \right) 
\frac{e^{{\rm H}(8 H^2 x/\Lambda^2)} - 1}{x} \left( x - \frac{1}{4} \right).  
\label{deSitterProp}
\ee
Notice that for the class of form factors we are considering here, ${\rm H}(z) = {\rm H}(-z)$. 
If $\Lambda_{\rm eff}$ is large with respect to $\Lambda$, we find three poles, see Fig.~\ref{figghost}. The second pole in the Fig.~\ref{figghost} corresponds to a ghost particle. 
The outcome of this analysis is a ghost instability of the approximate black hole solution. However,
in a non-local theory the instability is not catastrophic and can be estimated \cite{vilenkin, Maggiore}. 
Let us to consider the vacuum decay (in our case the black hole spacetime or actually the de Sitter spacetime) into a ghost particle and two normal gravitons \cite{khoury, vilenkin, Maggiore}, ${\rm BH} \rightarrow g, h,h$.
The decay probability per unit of volume and unit of time reads 
\be
\Gamma_{{\rm BH} \rightarrow g, h,h} = \frac{w}{V T} =  
\frac{\Lambda^6}{M_{\rm Pl}^2} e^{- {\rm H} (8 H^2 x_0/\Lambda^2)} \, , 
\ee
where $x_0$ is the ghost-like root in Fig.~\ref{figghost} and is obtained expanding the action near the ghost-pole. For the case of simplicity, here we assume 
$\Lambda = M_{\rm Pl}$. Therefore the lifetime is
\be
\tau_{{\rm BH} \rightarrow g, h,h} = \frac{1}{\Gamma_{{\rm BH} \rightarrow g, h,h} V} =  \frac{1}{V M_{\rm Pl}^4} e^{ {\rm H}(8 H^2 x_0/\Lambda^2)} \, .
\label{tau1}
\ee
The above decaying time is
finite and actually very long because the effective cosmological constant is proportional to the mass of the black hole~\cite{vilenkin}, namley 
\be
\tau_{{\rm BH} \rightarrow g, h,h} =  \frac{1}{V M_{\rm Pl}^4} e^{ {\rm H}( 8 M x_0/M_{\rm Pl})} \, .
\label{tau2}
\ee
If we consider an astrophysical object, $M$ is of order the Solar mass or more. The result is that the lifetime the all the processes of collapse, bounce and explosion take a very long time. 
The same exponential factor
can be inferred from the ghost-instability presented in~\cite{khoury},
replacing the Lorentz-violating scale 
with the scale of non-locality in the theory (\ref{theory}).

We now explicitly consider a class of form factors compatible with super-renormalizability and 
asymptotic polynomiality, namely 
\be
e^{{\rm H}(z)} = e^{\frac{1}{2} \left( \gamma_E + \Gamma(0, z^{2(\gamma+1)}) + \log z^{2(\gamma+1)} \right)} ,
\ee
whereby the decay time in the large mass limit simplifies to
 \be
\tau_{{\rm BH} \rightarrow g, h,h} \propto \frac{1}{V M_{\rm Pl}^4}
 \left( \frac{M}{M_{\rm Pl}} \right)^{\gamma+1}  
\ee
Taking $V=1/M_{\rm Pl}$ and $\gamma =2$, we exactly reproduces the Hawking result
\be
\tau_{{\rm BH} \rightarrow g, h,h} \propto \frac{1}{M_{\rm Pl}}
 \left( \frac{M}{M_{\rm Pl}} \right)^{3}   \, .
 \label{tau2}
\ee
It is quite impressive that the minimal super-renormalizable theory (the one for $\gamma =2$) 
embodies the Hawking evaporation process through the instability of the vacuum. 

Summarizing this section, we have shown that in a large class of weakly non-local gravitational theories 
any (approximate) black hole solution presenting a de Sitter core near $r=0$ is unstable due to the presence of a ghost instability. However, in these theories this is not a catastrophe because 
of the non-locality scale.
Therefore, 
the collapse of a cloud always produces a black supernova and never ends up with a black hole. 
Moreover, for the simplest range of theories compatible with super-renormalizability, the bouncing time perfectly agrees with the Hawking evaporation time. Despite this feature is not universal, it is 
impressive that it is a distinction of the minimal theory consistent at the quantum level.

\section{Conclusions \label{s-c}}

In Ref.~\cite{noi0,noi2}, we studied the gravitational collapse of a spherically
symmetric cloud in a class of weakly non-local theories of gravity that are a field theory proposal for a consistent theory of quantum gravity \cite{kuzmin, Tombo, modesto, modestoLeslaw}. 
However, in \cite{noi0,noi2} we only derived an approximate solution for the interior, while the external  
spacetime was completely conjectured, as we were not able to find a metric for the whole
spacetime. 
Nevertheless, we found a new picture for the gravitational collapse with the classical
singularity replaced by a bounce, after which the collapsing body starts
expanding. 
We inferred that black holes -- in the mathematical sense of regions covered by 
an event horizon -- do not form. The collapse only creates a temporary trapped surface, which can be interpreted as an event horizon only for a timescale shorter than the whole physical process. 
However, the latter might be extremely long for a stellar-mass object observed by a far-away observer. 
Our result is in agreement with those of other groups obtained with different
approaches~\cite{frolov,Frolov+-1,Frolov+,rovelli,torres,torres2,garay}.

In this paper, we have adopted a different approach to get an approximate solution for the
whole spacetime. Following the idea in~\cite{torres}, we have started from the
exterior region and assumed that the spacetime is static outside the matter. This is possible in
classical general relativity as a consequence of the Birkhoff theorem, and it
may be correct here as well. 
Such an assumption seems to
play a crucial role in the final fate of the collapse.

The approximate vacuum solution has two universal features: the spacetime near $r=0$ is well approximated by the de Sitter metric and the global structure show up an event horizon as well as a Cauchy internal horizon. If the mass is comparable to the Planck mass, there are no horizons at all.
It is clear that in a dynamical evolution of the black hole the Cauchy horizon instability is not a problem 
because it is just the internal part of o globally simply-connected trapped surface. 
These black holes are just like photo shoots of a non static but evolving black hole
(where by evolution we mean the dynamics of the black hole mass).

After imposing the boundary conditions, we have reconstructed the interior matter metric that,
in contrast to previous results reminded in the first part of the paper, 
does not show the expected bounce. On the contrary, there is an event horizon and a black hole does form. 
However, we have proved that the exterior metric is actually unstable due to the presence of a ghost-like
pole in the propagator. The instability here is not catastrophic because of the non-locality scale that actually allowed us to estimate the lifetime of the system (\ref{tau1}). 
It is quite remarkable that for the minimal super-renormalizable theory, the black hole lifetime
is identical to the Hawking evaporation time (\ref{tau2}).

\begin{acknowledgments}
C.B. acknowledges support from the NSFC grants No.~11305038 and No.~U1531117, the Thousand Young Talents Program, and the Alexander von Humboldt Foundation.
\end{acknowledgments}

\end{document}